\documentclass[12pt,preprint]{aastex}

\slugcomment{Submitted to ApJL}

\shorttitle{I~Zw~18 Distance and Age}
\shortauthors{Aloisi et al.}

\begin{document}


\title{I~Zw~18 revisited with {\it HST/ACS} and Cepheids:\\
New Distance and Age$^1$}


\author{A.~Aloisi\altaffilmark{2,3}, G.~Clementini\altaffilmark{4}, 
M.~Tosi\altaffilmark{4}, F.~Annibali\altaffilmark{2}, 
R.~Contreras\altaffilmark{4}, 
G.~Fiorentino\altaffilmark{4},
J.~Mack\altaffilmark{2}, 
M.~Marconi\altaffilmark{5}, I.~Musella\altaffilmark{5}, 
A.~Saha\altaffilmark{6}, M.~Sirianni\altaffilmark{2,3}, 
and R.~P.~van der Marel\altaffilmark{2}}


\altaffiltext{1}{Based on observations with the NASA/ESA Hubble
Space Telescope, obtained at the Space Telescope Science Institute,
which is operated by AURA for NASA under contract NAS5-26555}

\altaffiltext{2}{Space Telescope Science Institute, 3700 San Martin Drive, 
Baltimore, MD 21218; aloisi@stsci.edu}

\altaffiltext{3}{On assignment from the Space Telescope Division of the 
European Space Agency}

\altaffiltext{4}{INAF-Osservatorio Astronomico di Bologna, Via Ranzani 1, 
I-40127 Bologna, Italy}

\altaffiltext{5}{INAF-Osservatorio Astronomico di Capodimonte, Via 
Moiariello 16, I-80131 Napoli, Italy}

\altaffiltext{6}{National Optical Astronomy Observatory, P.O. Box 26732, 
Tucson, AZ 85726}


\begin{abstract}
We present new $V$ and $I$-band HST/ACS photometry of I~Zw~18, the
most metal-poor blue compact dwarf (BCD) galaxy in the nearby
universe.  It has been argued in the past that I~Zw~18 is a very young
system that started forming stars only $\lesssim 500$ Myr ago, but
other work has hinted that older ($\gtrsim 1$ Gyr) red giant branch
(RGB) stars may also exist. Our new data, once combined with archival
HST/ACS data, provide a deep and uncontaminated optical
color-magnitude diagram (CMD) that now strongly indicates an RGB. The
RGB tip (TRGB) magnitude yields a distance modulus $(m-M)_0 = 31.30
\pm 0.17$, i.e., $D = 18.2 \pm 1.5$ Mpc. The time-series nature of our
observations allows us to also detect and characterize for the first
time three classical Cepheids in I~Zw~18. The time-averaged Cepheid
$\langle V \rangle $ and $\langle I \rangle$ magnitudes are compared
to the $VI$ reddening-free Wesenheit relation predicted from new
non-linear pulsation models specifically calculated at the metallicity
of I~Zw~18. For the one {\it bona-fide} classical Cepheid with a
period of 8.63 days this implies a distance modulus $(m-M)_0 = 31.42
\pm 0.26$. The other two Cepheids have unusually long periods (125.0
and 129.8 d) but are consistent with this distance. The coherent
picture that emerges is that I~Zw~18 is older and farther away than
previously believed. This rules out the possibility that I~Zw~18 is a
truly primordial galaxy formed recently ($z \lesssim 0.1$) in the
local universe.\looseness=-2
\end{abstract}





\section{Introduction}
The BCD galaxy I~Zw~18 is one of the most intriguing nearby
objects. With $12 + \log (O/H) = 7.2$, corresponding to 1/50
Z$_{\odot}$ (Skillman \& Kennicutt 1993), it has the lowest nebular
oxygen abundance of all known star-forming galaxies in the nearby
Universe. It has a high gas fraction (e.g., van Zee et al.~1998) and
extremely high star formation (SF) rate per unit mass (Searle \&
Sargent 1972), producing a blue young stellar population that
dominates the integrated luminosity and color. All these observational
evidences make I~Zw~18 a chemically unevolved stellar system. As such,
it has long been regarded as a possible example of a galaxy undergoing
its first burst of star formation, hence a local analog to primordial
galaxies in the distant Universe.\looseness=-2

Many Hubble Space Telescope (HST) studies have focused on the
evolutionary state of I~Zw~18.  After other groups had already
resolved the brightest individual stars in the galaxy, our group
managed to detect fainter asymptotic giant branch (AGB) stars in
HST/WFPC2 images with ages of at least several hundreds Myr (Aloisi,
Tosi, \& Greggio 1999). These results were confirmed by \"Ostlin
(2000) through deep HST/NICMOS imaging. More recently, Izotov \& Thuan
(2004; IT04) presented new deep HST/ACS imaging observations.  Their
$I$ vs.~$V-I$ CMD shows no sign of an RGB (i.e., low-mass stars with
ages $\sim 1$-13 Gyr that are burning H in a shell around a He core)
at an assumed distance $D \lesssim 15$ Mpc.  Their conclusion is that
the most evolved (AGB) stars are not older than 500 Myr and that
I~Zw~18 is a {\it bona-fide} young galaxy.  This result was subsequently
challenged by Momany et al.~(2005) and our group (Tosi et al.~2006)
based on a better photometric analysis of the same data. This showed
that many red sources do exist at the expected position of an RGB, and
that their density in the CMD drops exactly where a RGB tip (TRGB, at
the luminosity of the He flash) would be expected. However, the small
number statistics, large photometric errors, and incompleteness, did
not allow a more conclusive statement.\looseness=-2

The actual nature of I~Zw~18 has important cosmological implications.  
According to hierarchical formation scenarios, dwarf ($M$ $\lesssim$ 
10$^9$ M$_{\odot}$) galaxies should have been the first systems to 
collapse and start forming stars. Indeed an RGB has been detected in 
all metal-poor dwarf irregular galaxies of the Local Group and BCDs 
within $D \lesssim 15$ Mpc that have been imaged with HST (e.g., 
SBS~1415+437, Aloisi et al.~2005, and references therein). I~Zw~18 
has remained the only elusive case so far. 

The lack of RGB evidence has also made it impossible to pinpoint the
distance of I~Zw~18 via the TRGB method. Its distance therefore
continues to be debated. With a recession velocity of $745 \pm 3$ km
s$^{-1}$, I~Zw~18 has often been assumed to be at a distance of $\sim
10$ Mpc ($H_0$ = 75 km\,s$^{-1}$Mpc$^{-1}$). Correction for
Virgocentric infall implies a slightly larger distance between 10 and
14.5 Mpc ($30.0 \lesssim m-M \lesssim 30.8$; \"Ostlin 2000). Izotov et
al.~(2000) argued that I~Zw~18 should be as distant as 20 Mpc to
provide consistency between the CMD, the presence of Wolf-Rayet stars
and the ionization state of the H~{\sc ii} regions. But they suggested
in IT04 a shorter distance $D \lesssim 15$ Mpc from the brightness of AGB
stars.\looseness=-2

In this Letter we present new $V$ and $I$ time-series photometry of
I~Zw~18 with HST/ACS that allows the first detection and study of the
galaxy's Cepheid stars. Our new data, once combined with the archival
ones, also yield a better CMD than has been hitherto available,
reaching about 1.5 mag below the TRGB. The results demonstrate that
I~Zw~18 is actually farther away and older than previously believed.

\section{Observations and Data Reduction}

Time-series photometry of I~Zw~18 was collected with ACS/WFC in 12
different epochs between 2005 October and 2006 January (GO program
10586, PI Aloisi). Due to the failure of the F814W exposure in one 
of the 12 visits (subsequently repeated), an additional epoch in 
F606W is available. 
Total integration times of $\sim 27700$ and $\sim
26200$ s were obtained in F606W and F814W, respectively. The 4
optimally dithered single exposures per epoch per filter were
registered by carefully measuring stars in the individual frames. The
pipeline-calibrated exposures were then corrected for geometric
distortion and co-added with cosmic-ray rejection using the
MultiDrizzle software to create an image per epoch each filter. The
images were drizzled onto a smaller pixel scale than the $0.05''$ ACS
pixels, to take advantage of the small scale dithers and improve the
$S/N$ for photometry. A similar procedure was used on all the single
exposures to create a master image in both filters.  Similarly, the
archival ACS/WFC data in F555W and F814W (GO program 9400, PI Thuan)
taken over a period of 11 days in 2003 May-June were combined into a
master image for a total integration time of $\sim 43500$ and $\sim
24300$ s in F555W and F814W, respectively. Single images per epoch
were also produced for the archival data (5 in F555W and 3 in 
F814W).\looseness=-2

The CMD shown in Figure~1 was obtained by performing PSF-fitting
photometry with the DAOPHOT/ALLSTAR package (Stetson 1987) on the
master images from our new and the archival data. Aperture corrections
were measured from stars in the frame and applied to all photometry.
Corrections for imperfect charge transfer efficiency (Riess \& Mack 
2004) resulted to be negligible (e.g., $\lesssim 0.03$ in all filters 
for a red TRGB star) and were not applied. Count rates were transformed 
to the Johnson-Cousins $V$ and $I$ magnitudes using the transformations 
given by Sirianni et al.~(2005), which have been shown to be accurate 
to $\sim 0.02$ mag. The values shown and discussed hereafter are
corrected for $E(B-V)=0.032$ mag of Galactic foreground extinction,
but not for any extinction intrinsic to I~Zw~18. For conciseness, we
refer to these magnitudes throughout this paper as $V$ and $I$ instead
of $V_0$ and $I_0$. The two ACS datasets were combined using several
different approaches and rejection schemes. The results discussed here
were obtained by demanding that stars should be detected in all the
four deep images ($V$ and $I$ for both datasets), yielding a catalog
of $\sim 2100$ stars. At the expense of some depth, this approach has
the advantage of minimizing the number of false detections and
therefore providing a relatively clean CMD. Annibali et al.~(2007, in
preparation) will present more details about the data reduction and
photometric analysis, as well as a Star Formation History (SFH)
analysis.\looseness=-2

PSF-fitting photometry of the single epochs in both our own and the
archival ACS/WFC time-series data was performed with the
DAOPHOTII/ALLSTAR/ALLFRAME package (Stetson 1987, 1994) to search for
and obtain light curves of the galaxy's variable stars. Candidate
variables were identified using four independent methods on the
single-epoch images: {\it i.)}  the Optimal Image Subtraction
Technique and the package ISIS 2.2 (Alard 2000), {\it ii.)} the
variability index (Welch \& Stetson 1993), {\it iii.)} an independent
PSF-fitting photometry performed with the DoPHOT program (Schechter,
Mateo, \& Saha 1993) to search for periodic variables as detailed in 
Saha \& Hoessel (1990), and {\it iv.)} visual inspection of a $\chi^2$ 
image, where each pixel is the $\chi^2$ value calculated from the
corresponding pixels in the co-registered target images at all epochs.
In addition, all suspected variables were examined by visually
`blinking' the relevant spots in the images at the various epochs.

The four procedures returned 4 confirmed variables. These are
discussed in Section~4 below and they are highlighted with open
circles in the CMD of Figure~1. In addition, some 30 other objects
were flagged as candidate variables, but we were generally unable to
find convincing periods nor otherwise classify them. Most of these
objects are in the main body of IZw18, where crowding complicates the
characterization of variables.

\section{The Color-Magnitude Diagram}

At magnitudes brighter than $I \simeq 27$ mag, we find in the CMD the
same star types previously detected with ACS by IT04: main-sequence
(MS) stars and their more evolved counterparts, i.e., young
supergiants (SGs) and AGB stars.  A ``blue plume'' which is the
combined feature of MS stars and blue SGs (evolved stars at the hot
edge of their core He-burning phase), shows up near $V-I \simeq 0.0$
mag. At $V-I \simeq 1.3$ mag, red SGs appear at magnitudes $I \lesssim
24$ mag, while AGB stars dominate in the magnitude range $I \simeq
25$--27 mag and form the upper part of what we will call a ``red plume''. 
The morphology of the AGB star features compares well with the positions 
at which oxygen-rich and carbon-rich AGB stars are predicted in 
evolutionary models (e.g., Figure~12 of Marigo, Girardi, \& Chiosi 2003). 
Our ACS data show that both the blue and red plumes extend down to much
fainter magnitudes. But more importantly, they reveal the presence of
a much older stellar population.  Indeed, below $I \simeq 27.3$ mag
and around $V-I \simeq 1.3$ mag, we detect faint red stars exactly at
the position where an RGB would be expected (see the Padua isochrones
overplotted in Figure~1). The small median error around $I \simeq
27.5$ mag suggests that these faint red stars are not more massive
younger stars with large photometric uncertainties.

Figure~2 shows the $I$-band luminosity function (LF) for stars in
I~Zw~18 CMD with red colors in the range $V-I = 0.75$--1.5 mag. This
LF presents a sharp drop towards brighter magnitudes, exactly as would
be expected from a TRGB. We used a Savitzky-Golay filtering technique
developed by one of us (R.~P.~v.~d.~M.) and described in Cioni et
al.~(2000) to determine the TRGB magnitude. This yields $I_{TRGB} =
27.27 \pm 0.14$ mag, with the error bar being dominated by random
uncertainties from the finite number of stars. At the metallicity of
I~Zw~18, the absolute magnitude of the TRGB discontinuity is
$M_{I,TRGB} = -4.03 \pm 0.10$ (Bellazzini et al.~2004). This implies a
distance modulus $(m-M)_0 = 31.30 \pm 0.17$ mag, i.e., $D = 18.2 \pm
1.5$ Mpc, assuming that the evolved RGB stars have negligible
intrinsic extinction (see, e.g., Cannon et al.~2002).

The evidence for an RGB in I~Zw~18 is further strengthened by
comparison to another BCD, SBS~1415+437, observed by us with a similar
HST/ACS set-up (Aloisi et al.~2005). This galaxy is not quite as metal
poor as I~Zw~18 ($12 + \log (O/H) = 7.6$) and is nearer at $D \simeq
13.6$ Mpc. But taking into account the differences in distance,
corresponding to a $\Delta (m-M) \simeq 0.61$ mag, the two BCDs have
very similar LFs (see Figure~2). Since SBS 1415+437 has an
unmistakable RGB sequence, this suggests that such an RGB sequence
exists in I~Zw~18 as well. The larger distance of I~Zw~18 explains why
with a similar observational setup we detect $\sim 10$ times fewer
stars than in SBS 1415+437.

\section{Classical Cepheids in I~Zw~18}

Periods and classification in type of the 4 confirmed variable stars
in I~Zw~18 were derived from the study of the $V$ and $I$ light curves
separately, using GRaTiS (Graphical Analyzer of Time Series), custom
software developed at the Bologna Observatory by P. Montegriffo 
(Clementini et al.~2000 and references therein) which uses both the Lomb 
periodogram and light-curve fitting with a truncated Fourier series.

Our ACS/WFC data along with the archival ones span a total time
baseline of about 2.5 years. The combined dataset allowed us to derive
reliable periods for three of the four confirmed variable stars. All
of them are classified as classical Cepheids, both based on the shape
of the light curve and the star's position on the CMD, where they all
fall within or close enough to the instability strip for Classical
Cepheids at the metallicity of I~Zw~18. One of these Cepheids has a
period of 8.63 d (see Figure~3), well within the typical range used to
statistically define period-luminosity (PL) relations in other
galaxies. The other two Cepheids have instead periods of 125.0 d and
129.8 d, respectively. This is much longer than ever observed in other
galaxies, with the periods of Cepheids typically being $P \lesssim
70$--80 d (see Udalski et al. 1999; Freedman et al. 2001). The
precision of our period determinations is of the order of 1-2 decimal
places. The residuals from the truncated Fourier series that best fit
the folded light curves are in the ranges 0.02--0.12 and 0.02--0.11
mag, in $V$ and $I$, respectively.\looseness=-2

The fourth confirmed variable is very bright and red. Only a portion
of its light curves is covered by our observations, leading to two
possible alternative periodicities of 139 and 186 days
respectively. In spite of the unusually long period and the red color,
probably due to the star being blended by a red companion, both the
high luminosity and the shape of the light curve suggest that this
variable star is probably a Classical Cepheid as well. In absence of a
well-determined period, it will not be considered further in our
discussion below. A more detailed description of the sample of
variable stars, and the data analysis performed to identify and
characterize them, will be presented in Fiorentino et al.~(2007, in
preparation). This sample will be unique for probing the properties of
variable stars at previously unexplored metallicities. Here we focus
merely on the implied distance of I~Zw~18.

New non-local time-dependent convective pulsating theoretical models
of Classical Cepheids were computed at the proper metallicity of
I~Zw~18 ($Z = 0.0004$, $Y=0.24$) by our group (Marconi et al.~2007, in
preparation). These models build on our previous modeling expertise at
higher metallicities (e.g., Marconi, Musella, \& Fiorentino 2005;
Fiorentino et al.~2007). The new models were used to obtain a
theoretical reddening-free $VI$ Wesenheit relation that was directly
compared to the Wesenheit relation of the observed Cepheids in order
to infer a distance. The 8.63 d Cepheid yields a distance modulus
$(m-M)_0 = 31.42 \pm 0.26$ mag if a canonical mass-luminosity relation
is assumed in the model computations and of $(m-M)_0 = 31.27 \pm 0.26$
mag in case of overluminous noncanonical models (e.g., Caputo et
al.~2005; Fiorentino et al.~2007). We have no models yet that
reproduce the very long-periods of the other two Cepheids with
well-determined periods, so the Wesenheit relation needs to be
extrapolated at such long periods to obtain a distance. But if we do
this, the average of all three Cepheids yields a very similar
distance, $(m-M)_0 = 31.38 \pm 0.17$.

The Cepheid distance estimates of I~Zw~18 are in excellent agreement
with our new TRGB distance. This agreement supports the view that the
feature identified in Figure~1 is indeed the TRGB, and therefore, that
I~Zw~18 does indeed have RGB stars. In turn, this implies that the
galaxy cannot have formed in the last $\sim 1$ Gyr. Moreover, the RGB
stars are unlikely to be as young as 1--2 Gyr, because for such ages
the TRGB is up to $\sim 2$ mag fainter than the standard candle value
used here that applies to populations with ages $\gtrsim 2$ Gyr (e.g.,
Figure~2 of Barker, Sarajedini, \& Harris 2004).

The extraordinary long periods of several of the I~Zw~18 classical
Cepheids deserve some further comments. They could be a metallicity
effect, since I~Zw~18 is the lowest metallicity galaxy where Cepheids
have been observed. They could also be related to the peculiar SFH of
this BCD, since I~Zw~18 is currently experiencing a strong starburst
that makes it more likely to detect more massive (i.e., brighter and
shorter-lived) and thus longer-period Cepheids than in more regular
(spiral) galaxies. Particularly striking is also the lack of observed
Cepheids with periods anywhere between $\sim 10$--120 days. We
attribute this absence to the lack of a detectable SF activity in
I~Zw~18 at the epochs when the missing Cepheids should have formed.
Indeed, once we scale the SFH derived by Aloisi et al.~(1999) from 10
Mpc to the new distance derived here, we expect few stars in the mass
range 6--20 M$_{\odot}$ (corresponding to pulsation periods of $\sim
10$--120 days) currently in the instability strip.

\section{Discussion and Conclusions}

We have obtained new deep HST/ACS observations of I~Zw~18 that provide
improved insight into the evolutionary state of this benchmark
metal-poor BCD. Our results indicate that this galaxy contains RGB
stars, in agreement with findings for other local metal-poor BCDs
studied with HST. Underlying old ($\gtrsim 1$ Gyr) populations are
therefore present in even the most metal-poor systems, so they must
have started forming stars at $z \gtrsim 0.1$. Deeper studies (well
below the TRGB) will be needed to pinpoint the exact onset of SF in
these galaxies. We also find that our TRGB distance of I~Zw~18, $D =
18.2 \pm 1.5$ Mpc, which is confirmed by our Cepheids results, places
the galaxy farther away than the values $\sim 15$ Mpc that have often
been assumed in previous work. This may explain why it has remained
difficult for so long to unambiguously detect or rule out the presence
of old resolved (RGB) stars in this galaxy.



\acknowledgments

Support for proposal \#10586 was provided by NASA through a grant from
STScI, which is operated by AURA, Inc., under NASA contract NAS
5-26555. GC and MT acknowledge support from PRIN-INAF 2005. We are
grateful to Marcella Maio for her invaluable contribution in setting
the parameters of the photometric ACS data reductions.\looseness=-2

\begin{figure}
\epsscale{0.55}
\plotone{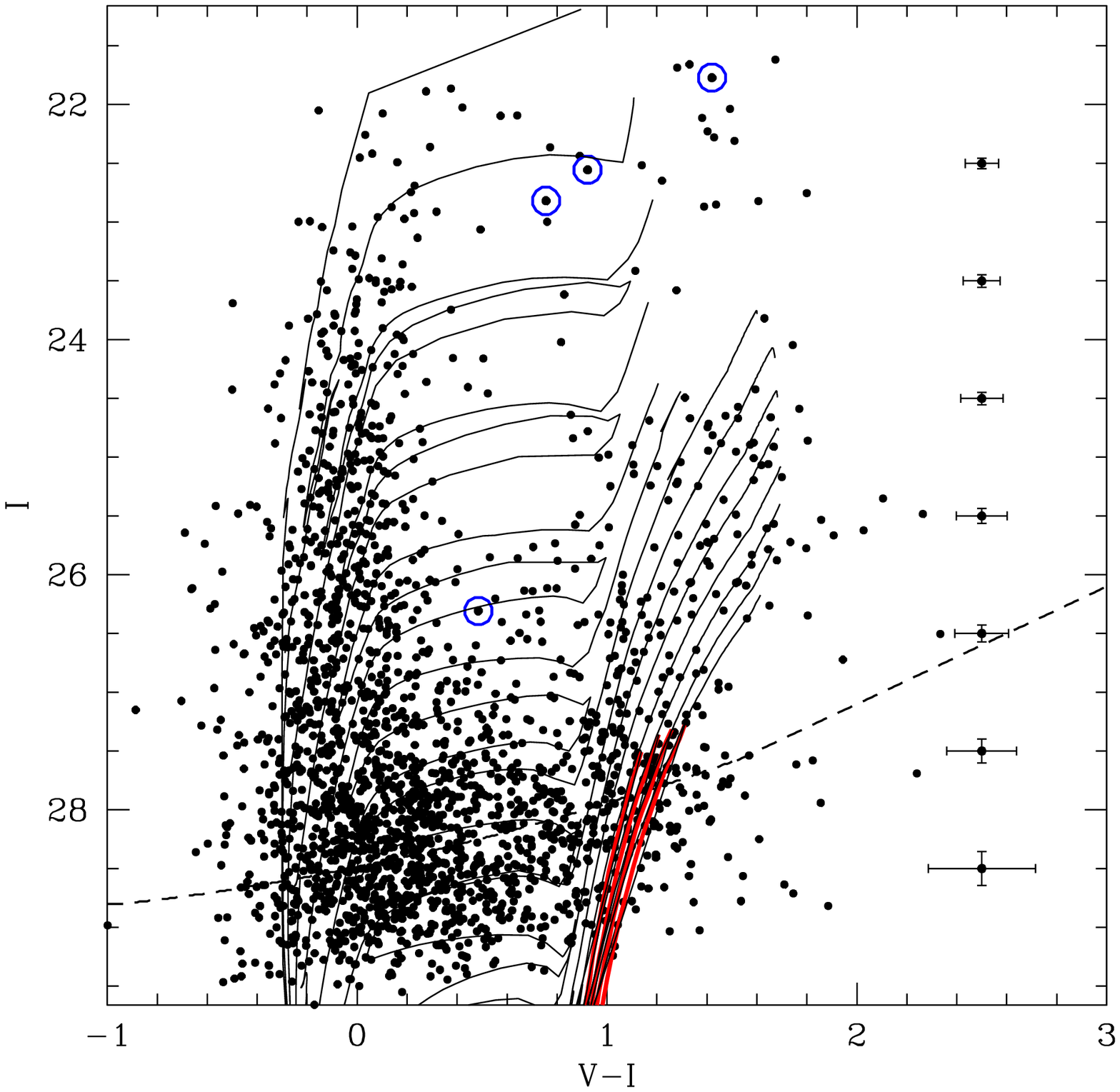}
\vspace{0.2cm}
\caption{Johnson-Cousins $I$, $V-I$ CMD of the resolved stellar
population in both the main and secondary bodies of I~Zw~18. The data
are corrected for foreground extinction, but not for possible
extinction internal to the galaxy. Padua isochrones for {\it (roughly
from top left to bottom right)} log (age) = 6.75, 7.00, 7.25, ...,
9.75, 10.00 are overlaid, with the RGB phase for isochrones with log
(age) = 9.25, 9.50, 9.75, and 10 colored red (Bertelli et
al.~1994). The isochrones have metallicity $Z = 0.0004$ (as inferred
from the H~{\sc ii} regions of I~Zw~18) and are shown for the distance
of $D=18.2$ Mpc ($m-M = 31.30$) inferred from the TRGB.  Blue open
circles highlight the four confirmed variables (see Section~4), which
are plotted according to their intensity-averaged magnitudes. Their
periods are (from bottom to top) 8.63d, 129.8d, and 125.0d. The period
of the brightest variable is even longer, but its exact value is not
well determined. Median photometric errors at $V-I = 1$ (determined by
comparison of measurements from GO-9400 and GO-10586) are shown as
function of $I$-band magnitude on the right. The dashed curve is an
estimate of the 50\% completeness level (obtained by fitting the
combination of a parametrized completeness function and a model Star
Formation History (SFH) to the observed CMD). More sophisticated
estimates of the errors and completeness based on artificial star
tests will be presented in Annibali et al.~(2007, in preparation).
\label{fig2}}
\end{figure}

\begin{figure}
\epsscale{1.0}
\plotone{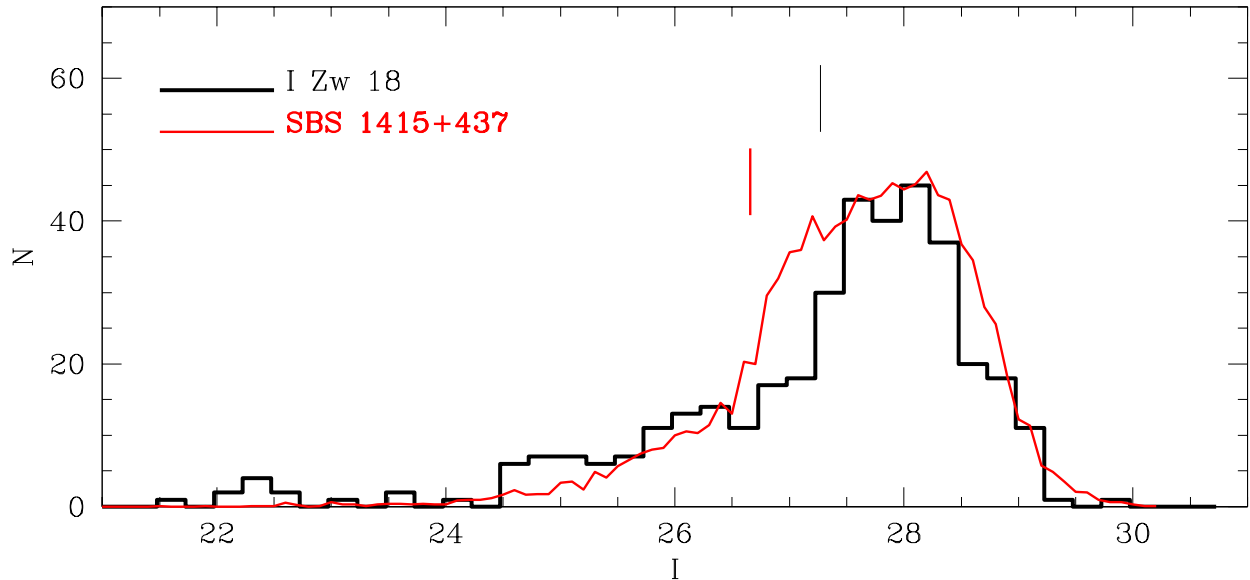}
\vspace{0.2cm}
\caption{$I$-band LF histogram for stars in I~Zw~18, showing (in
black) the number of stars $N$ with colors in the range $V-I =
0.75$--1.5 mag per $0.25$ mag $I$-magnitude bin, inferred from the CMD
in Figure~1.  For comparison, the LF of SBS~1415+437 from the ACS data
presented in Aloisi et al.~(2005) is also plotted (in red),
arbitrarily renormalized. Vertical marks indicate the positions of the
TRGB, determined as described in the text. At these magnitudes there
is a steep LF drop towards brighter magnitudes, due to the end of the
RGB sequence. By contrast, the LF drop towards fainter magnitudes at
$I \gtrsim 28$ mag is due to incompleteness in both cases. Apart from
a distance shift $\Delta (m-M) \simeq 0.61$, these metal-poor BCD
galaxies have very similar LFs.
\label{fig3}}
\end{figure}

\begin{figure}
\epsscale{1.00}
\plotone{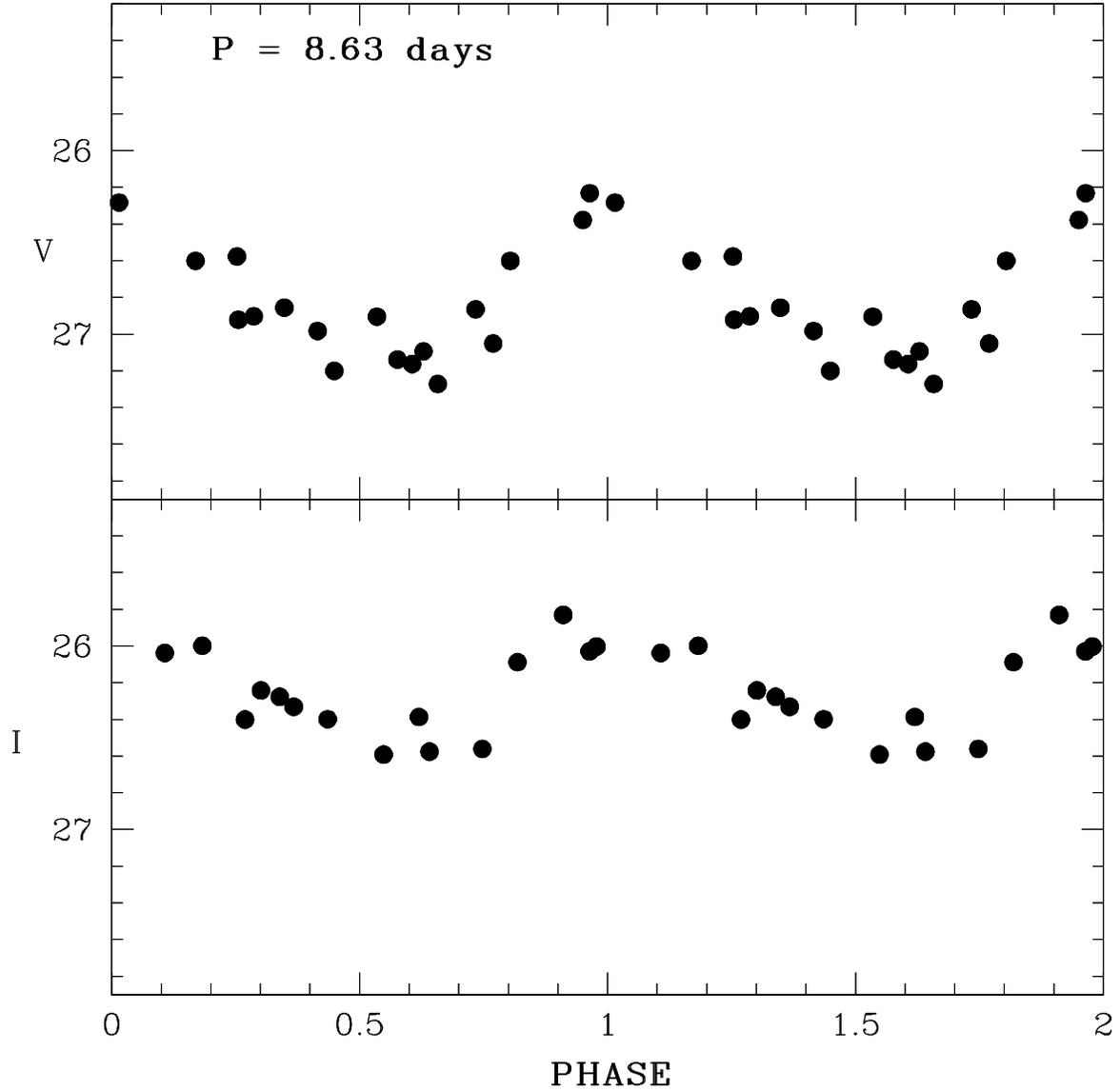}
\vspace{0.2cm}
\caption{$V$ and $I$ light curves, corrected for Galactic foreground
extinction, for the 8.63 d Cepheid. The single-epoch photometry was
calibrated to the Johnson-Cousins $V$ and $I$-bands by calculating the
zero point offset between the intensity-averaged instrumental
magnitudes of the light curves and the DAOPHOT/ALLSTAR photometry of
the master images.
\label{fig4}}
\end{figure}

\end{document}